\documentclass[twoside,english,nofootinbib,preprintnumbers,aps,11pt,showpacs]{revtex4}

\usepackage{euscript,amssymb}
\usepackage{amsthm}
\usepackage{graphicx}
\usepackage{amsfonts}
\usepackage{amsmath}
\usepackage{amssymb}
\usepackage{fancyhdr}
\usepackage{epsfig}
\usepackage{color}
\usepackage{esint}
\usepackage{soul}
\usepackage{afterpage}
\usepackage[papersize={8.5in,11in}]{geometry}

\newcommand{\be}{\begin{equation}}
\newcommand{\ee}{\end{equation}}

\begin{document}

\title{Spacetimes foliated by non-expanding and Killing horizons: higher dimension}

\author{Jerzy Lewandowski}
\email[]{Jerzy.Lewandowski@fuw.edu.pl}
\author{Adam Szereszewski}
\email[]{Adam.Szereszewski@fuw.edu.pl}
\author{Piotr Waluk}
\email[]{Piotr.Waluk@fuw.edu.pl}
 
\affiliation{\vspace{6pt} Institute of Theoretical Physics, Faculty of
  Physics, University of Warsaw, Pasteura 5, 02-093 Warsaw, Poland}
\begin{abstract}  The theory of non-expanding horizons (NEH) geometry and the theory of near horizon geometries (NHG)  are two mathematical relativity 
frameworks generalizing the black hole theory. From the point of view of the NEHs theory, a NHG is just a very special case of a spacetime containing an NEH of many extra symmetries.  It can be obtained as the Horowitz limit of a neighborhood of an arbitrary extremal Killing horizon.       
An unexpected relation between the two  of them, was discovered in the study of spacetimes foliated by a family of NEHs.  The class of 
4-dimensional NHG solutions (either vacuum or coupled to a Maxwell field) was found  as a family of examples of spacetimes admitting a NEH foliation.  
In the current paper we systematically investigate  geometries of the NEHs foliating a spacetime for arbitrary  
matter content and in arbitrary  spacetime dimension. We find that each horizon belonging to the foliation satisfies a condition that may be interpreted as an invitation for a transversal extremal Killing horizon to exist.  Assuming the existence of a transversal  extremal Killing horizon, we derive all the spacetime metrics satisfying the vacuum Einstein's equations.                    
\end{abstract}



\maketitle

\section{Introduction}
The theory of non-expanding horizons (NEH) geometry \cite{ABL,LivRevIH,LPhigh} and the theory of near horizon geometries (NHG) \cite{Horowitz,Reall,LivRevNHG} are two mathematical relativity frameworks generalizing the black hole theory. From the point of view of the NEHs theory, a NHG is just a very special case of a spacetime containing an NEH of many extra symmetries.  It can be obtained as the Horowitz limit of a neighborhood of an arbitrary extremal Killing horizon.       
Another, unexpected relation between the two  of them, was discovered in the study of spacetimes that can be foliated by a family of NEHs \cite{PLJ}. Spacetimes foliated by non-expanding horizons are also known as Kundt's 
class of spacetimes \cite{ExactSolutions}.  The NEH context, however, brings new geometric ideas. The class of 
4-dimensional NHG solutions (either vacuum or coupled to a Maxwell field) was found in \cite{PLJ} as a family of examples of spacetimes admitting a NEH foliation.  We generalize that result in two directions.  The first one is a 
systematic investigation of geometries of the NEHs foliating a spacetime valid for arbitrary  
matter content of the foliated spacetime. The second is passing from the $4$ to an arbitrary number of spacetime dimension. We find that each horizon belonging to the foliation satisfies a condition that may be interpreted as an invitation for a transversal extremal Killing horizon to exist.  On the other hand, assuming the existence of a 
transversal  extremal Killing horizon, we derive all the spacetime metrics satisfying the vacuum Einstein equations.

\section{Non-expanding and isolated horizons}
\subsubsection{Notation}
 We consider in this paper an $n$ dimensional manifold $M$.  In $M$ we will consider co-dimension $1$ 
and co-dimension $2$ surfaces. It will be convenient to use the following index convention:  
\begin{itemize}
\item   $X^\alpha$, $Y_\beta$ for vectors and, respectively, co-vectors in $M$,
\item  $X^a$, $Y_a$ for vectors and, respectively covectors  in a co-dimension $1$ surface
\item $X^A$, $Y_A$ for vectors and, respectively  covectors in a co-dimension $2$ surface 
\item moreover, given a covector $Z_{\alpha}$, its pullback to a given surface of co-dimension
\begin{itemize}
\item  $1$  will be denoted by $Z_a$
\item  $2$  will be denoted by $Z_A$
\end{itemize}
\end{itemize}
Notice, that a vector $X^A$ can  be still consistently denoted by $X^a$, and a vector $Y^a$ can be denoted
by $Y^\alpha$.  

The manifold $M$ is endowed with a spacetime metric tensor $g_{\alpha\beta}$ and the signature convention we use is
$-+\,...\,+$. 

\subsubsection{Definition of non-expanding horizon (NEH)}
A non-expanding horizon in $M$ is, briefly speaking,  a non-expanding null  surface 
 of co-dimension 1 which admits compact spacelike slices.  More precisely, 
 a non-expanding horizon is a co-dimension 1 surface
     $$H\ \subset\ M$$ 
such that:   %
\begin{itemize}
\item the induced  in $H$ metric $g_{ab}$  satisfies:
\begin{itemize}
\item at each point of $H$ there is a non-trivial degenerate vector $\ell^a$, that is
\be\label{l} \ell^a g_{ab}\ =\ 0\ee
\item a degenerate vector field $\ell$ satisfies
\be\label{Llg} {\cal L}_\ell g_{ab}\ =\ 0 \ee
\end{itemize}
\item $H$ may be  foliated by compact, spacelike co-dimension 2 surfaces. 
\end{itemize}
\subsubsection{Einstein's equations, energy conditions}
On  non-expanding horizons we are assuming Einstein's equations
\be\label{Ein} G_{\alpha\beta}\ =\  8\pi T_{\alpha\beta}\ee
and the following energy inequalities
\be\label{energy}   T_{\ell\ell} \ \ge \ 0,\ \ \ \ \ \ \ \ \ \ \   T_{\ell\alpha}{T_\ell}^\alpha\ \le\ 0 .\ee
The conditions (\ref{l}), (\ref{Llg}) and (\ref{energy}) are invariant
with respect to transformations  
$$ \ell\ \mapsto\ f\ell ,$$
where $f$ is an arbitrary function defined on $H$. The condition (\ref{Llg}) is often replaced
by the  weaker assumption, that the expansion of $\ell$ vanishes (while the shear may be arbitrary). 
However, then,  (\ref{Llg})  follows from  the $n$-dimensional generalization of the Raychaudhuri equation \cite{LPhigh}
due to (\ref{energy}) (that is, eventually, the shear of $\ell$ is forced to vanish). 

\subsubsection{Geometry of NEH}
The spacetime covariant derivative $\nabla_\alpha$ reduces to  the tangent bundle $T(H)$ in the sense that
for every two vector fields $X^\alpha$ and $Y^\beta$ in $M$ tangent to $H$, the vector field
\be Y^\alpha\nabla_\alpha X^\beta \ee
is again tangent to $H$. The inducted derivative $\nabla_a$ still satisfies
\be  \nabla_ag_{bc}\ =\ 0 ,\ee
and is torsion free, however, it is not determined by $g_{ab}$. The pair $(g_{ab},\nabla_c)$ is intrinsic-extrinsic 
geometry of  a non-expanding horizon.    

\subsubsection{Rotation, surface gravity and the $0$th law.}
A rotation 1-form potential of a non-expanding horizon $(H, g_{ab}, \nabla_a)$  is a 1-form $\omega_a$
on $H$ defined by a null vector field $\ell^b$ on $H$ and the equality  
\be \nabla_a\ell^b\ =\ \omega^{\ell}_a\ell^b .\ee
Upon the transformations
\be
   \label{lmapstofl} \ell \mapsto f\ell, \qquad f=f(x^a)
\ee
the rotation 1-form potential transforms as follows
\be  \omega^{f\ell}_a\ =\ \omega^{\ell}_a + \nabla_a {\rm ln}f . \ee
The rotation invariant is
\be \Omega_{ab}\ :=\ \nabla_{a}\omega^\ell_{b} \ -\ \nabla_{b}\omega^\ell_{a}, \ee
while the surface gravity $\kappa^{\ell}$  is defined as
 \be\label{kappa} \kappa^{\ell}\ =\ \ell^a\omega^{\ell}_a . \ee 
 
 Due to Einstein's equations and the energy inequalities (\ref{energy}) the surface gravity $\kappa^\ell$ and 
 rotation 1-form potential $\omega^\ell_a$  satisfy a constraint
 \be\label{dkappa}  \nabla_a \kappa^{\ell}\ =\ {\cal L}_\ell \omega^{\ell}_a .\ee  

 \subsubsection{Isolated horizons}\label{sec:iso}
An infinitesimal symmetry of a non-expanding horizon $H$ is a vector field $X$ tangent to $H$ and such
that 
\be {\cal L}_X g_{ab}\ =\ 0\ =\ [{\cal L}_X, \nabla_a]. \ee

Isolated horizon is a non-expanding horizon which admits a null infinitesimal symmetry. 
That is, $H$ is an isolated horizon, whenever we can choose a degenerate (but non-trivial) 
vector field $\ell$  in $H$ such that in addition to  (\ref{Llg}), 
\be\label{iso} [{\cal L}_\ell, \nabla_a]\ =\ 0 . \ee
In general, the condition is not invariant with respect to the transformations (\ref{lmapstofl}),
except for a constant non-vanishing $f=f_0$ and some very special cases \cite{ABL}.

At every isolated horizon  
\be {\cal L}_\ell \omega^\ell_a\ =\ 0, \ee
hence  the constraint  (\ref{dkappa}) implies
\be \kappa^\ell\ =\ {\rm const} .\ee
 By the analogy to the black hole termodynamics it is called the $0$th law of isolated horizons 
 thermodynamics \cite{ABL,LPhigh}. 

An isolated horizon $(H, g_{ab},\nabla_a,\ell^a)$ is called extremal, whenever 
\be \kappa^\ell\ =\ 0 \ee
for the null symmetry  $\ell$.   
That property is invariant with respect to the rescaling  by a constant
$$\ell\mapsto f_0\ell.$$
There exists however an isolated horizon which admits two
linearly independent null symmetries, and which is extremal with respect to the first one, and  non-extremal
with respect to the second \cite{ABL} (it will appear in the current paper later). That is why in the definition of the extremality
we also declare the generator $\ell$.  

\subsubsection{Einstein's  constraints at NEH - a covariant form}
For every non-expanding horizon the intrinsic-extrinsic geometry $(g_{ab}, \nabla_a)$ satisfies 
constraints implied by  Einstein's equations and the energy inequalities (\ref{energy}) \cite{LPhigh}.
The first constraint has been already mentioned (\ref{dkappa}). 

To spell out the remaining constraints we introduce a coordinate $v$ on $H$ such that
\be\label{v} \ell^a\nabla_av\ =\ 1,\ee 
and consider
\be  S_{ab}\ := -\nabla_a\nabla_b v .\ee
Then, the relation between  $(g_{ab}, \nabla_a)$   with the spacetime Riemann tensor $R^\alpha{}_{\beta\gamma\delta}$ 
is
\be\label{LlSab}
{\cal L}_\ell S_{ab}\ =\ \nabla_{(a}\omega^\ell_{b)}\ +\  \omega^\ell_{a}\omega^\ell_{b} \ -\ R_{c(ab)}{}^d\ell^c\nabla_dv.
\ee


\subsubsection{Einstein's constraints at NEH - the longitudinal and transversal parts}
Since,
\be \ell^a S_{ab}\ =\ \omega^\ell_b,\ee
the contraction of $\ell^a$ with (\ref{LlSab}) gives the constraint (\ref{dkappa}) we have already 
invoked.

A  2-dimensional slice $S_{v_0}$ of $H$ defined by all the points such
that 
 \be  v(x^a)\ =\ v_0,\ee
 is equipped with  the induced metric $g_{AB}$,  its covariant derivative  $\nabla^{(n-2)}_A$ (metric, torsionfree),
and the corresponding Ricci tensor $R^{(n-2)}_{AB}$, as well as the pullback $R_{AB}$ of the spacetime
Ricci tensor.  Consider the foliation of $H$ by all the slices $S_v$. 
 
The pullback of (\ref{LlSab}) to each slice  amounts to
\be\label{LlSAB}
{\cal L}_\ell S_{AB}\ =\ -\kappa^\ell S_{AB}\ + \nabla^{(n-2)}_{(A}\omega^\ell_{B)}\ +\  \omega^\ell_{A}\omega^\ell_{B} \ -\ \frac{1}{2}R^{(n-2)}_{AB} \ +\ \frac{1}{2}R_{AB} .
\ee

\subsubsection{Constraints on the Riemann and Ricci tensors at NEH}
The energy momentum tensor $T_{\alpha\beta}$ determines the pullback $R_{AB}$ onto $S_v$ of the spacetime
Ricci tensor $R_{\alpha\beta}$ and in this way constraints the horizon geometry.  On the other hand, $T_{\alpha\beta}$ is 
constrained itself by the identities satisfied at $H$ \cite{LPhigh} by the spacetime Ricci and respectively Riemann tensor 
$R^\alpha{}_{\beta\gamma\delta}$,
\be\label{RatH} R_{a\beta}\ell^\beta\ =\ 0\ =\ R_{abc\delta} \ell^\delta. \ee

\subsubsection{Einstein's constraints at isolated horizons} 
 If $(H, g_{ab}, \nabla_a,\ell^a)$ is an isolated  horizon, then 
 \be {\cal L}_\ell S_{AB}\ =\ 0 ,  \ee
hence at every slice $S_v$, 
\be\label{Isoconstr}
-\kappa^\ell S_{AB}\ + \nabla^{(n-2)}_{(A}\omega^\ell_{B)}\ +\  \omega^\ell_{A}\omega^\ell_{B} \ -\ 
\frac{1}{2}R^{(n-2)}_{AB} \ +\ \frac{1}{2}R_{AB} \ =\ 0.
\ee 
%
It follows, that in addition to (\ref{RatH}) at every isolated horizon
such that (\ref{energy}) we have
\be    
{\cal L}_\ell R_{a b}\ =\ 0 . \ee 

If $(H, g_{ab}, \nabla_a,\ell^a)$ is an extremal isolated  horizon, then at every slice $S_v$,
\be\label{Iso1}
 \nabla^{(n-2)}_{(A}\omega^\ell_{B)}\ +\  \omega^\ell_{A}\omega^\ell_{B} \ -\ 
\frac{1}{2}R^{(n-2)}_{AB} \ +\ \frac{1}{2}R_{AB}\ =\ 0 . 
\ee

\section{Existence of isolated horizon structure at a NEH}

Before formulating  an inverse proposition, notice that given a non-expanding horizon $H$ 
and a tangent null vector $\ell$,  we do not a priori know whether $H$ has a null symmetry that makes 
him isolated in the sense of (\ref{iso}) or not,  and if it has,  what choice of a null vector field 
\be\label{fell}\ell'=f\ell\ee  
provides  the actual generator such that
\be [{\cal L}_{\ell'},\nabla_a]\ =\ 0 . \ee
This issue was intensively studied in \cite{ABL}, and a necessary and sufficient condition was derived.  
Here, we formulate and prove another  sufficient condition that will be  applied  in the next section.

 
Suppose $H$ is a non-expanding horizon and $\ell$ is a null vector field
tangent to $H$ such that the following conditions are satisfied 
\begin{itemize}
\item on $H$
\be\label{Iso2} {\cal L}_{\ell} R_{a b}\ =\ 0  ,\ee 
\item there is  a spacelike slice $S$ of $H$,  such that
\be \kappa^\ell|_S\ =\ {\rm const} ,\ee
and 
\be\label{onS} 
\left.\left(-\kappa^\ell S_{AB}\ + \nabla^{(n-2)}_{(A}\omega^\ell_{B)}\ +\  \omega^\ell_{A}\omega^\ell_{B} \ 
-\ \frac{1}{2}R^{(n-2)}_{AB} \ +\ \frac{1}{2}R_{AB}\right)\right|_S=\ 0  ,\ee  
where $S_{AB}$ is defined by the function $v$ constant at $S$ and such that $\ell^a\nabla_a v=1.$
\end{itemize}
Then, there is a null vector field $\ell'$ (\ref{fell}) on $H$ such that
\begin{align}\label{conc} 
\kappa^{\ell'}\ &=\ {\rm const} \nonumber\\
 [{\cal L}_{\ell'},\nabla_a]\ =\ 0,\ \ \ &{\rm and} \ \ \ \left.\left(\kappa^{\ell'} - \kappa^{\ell}\right)\right|_S \ =\ 0
 \end{align}
In a matter of fact, the bullet above is a condition on $R_{ab}$ rather than on $\ell$ - once it holds, it continues
to hold for any other $\ell'$ (\ref{fell}).               
     
To prove the theorem we just indicate the correct $\ell'$.  Given the vector field $\ell$ and the corresponding 
$\kappa^\ell$ which is not necessarily constant on all $H$, there always exists another null vector field $\ell'$, 
such that on $S$
\be \ell'|_S\ =\ \ell|_S, \ee
and on $H$
\be\label{kappakappa'}  \kappa^{\ell'}\ =\ \kappa^\ell|_S . \ee
We will demonstrate now, that this vector field $\ell'$ satisfies the conclusion (\ref{conc}).
It already does satisfy the second equality of  (\ref{conc}). To prove the first one,
it is sufficient to show
\be\label{LomLS}{\cal L}_{\ell'}\omega^{\ell'}_a\ =\  0\ =\ {\cal L}_{\ell'}S'_{AB} .\ee
But the assumption (\ref{kappakappa'}) together with the zeroth low (\ref{dkappa}) imply the first equation
above. The second equation will take a few steps.           
First, we find the corresponding  $\omega^{\ell'}_A|_S$,
 \be\label{omega'onS}  \omega^{\ell'}_A|_S\ =\ \omega^\ell_A |_S\  + \nabla_A {\rm ln}f |_S\  =\  
 \omega^\ell_A |_S \ee
because                       
\be  f|_S \ =\ 1 .\ee
From that we conclude the following equality which holds on the slice $S$
\begin{multline}
\label{onS'} \left.\left(\nabla^{(n-2)}_{(A}\omega^{\ell'}_{B)}\ +\  \omega^{\ell'}_{A}\omega^{\ell'}_{B} \ -\ 
\frac{1}{2}R^{(n-2)}_{AB} \ +\ \frac{1}{2}R_{AB}\right)\right|_S \ =\\
  = \left.\left(\nabla^{(n-2)}_{(A}\omega^{\ell}_{B)}\ +\  
\omega^{\ell}_{A}\omega^{\ell}_{B} \ -\ 
\frac{1}{2}R^{(n-2)}_{AB} \ +\ \frac{1}{2}R_{AB}\right)\right|_S 
\end{multline}
where we added the $f$ independent terms to suggest the relation with (\ref{LlSAB}). 
But due to the first equality in (\ref{LomLS}), and the equality (\ref{Iso2})  
\be\label{onH} {\cal L}_{\ell'} \left(\nabla^{(n-2)}_{(A}\omega^{\ell'}_{B)}\ +\  \omega^{\ell'}_{A}\omega^{\ell'}_{B} \ -\ 
\frac{1}{2}R^{(n-2)}_{AB} \ +\ \frac{1}{2}R_{AB}\right)\ =\ 0   \ee
on $H$. 

The remaining part of the proof depends on whether $\kappa^{\ell'}$ vanishes or not. Consider first the extremal case, that is 
\be \kappa^{\ell'}\ =\ 0.\ee
It follows from (\ref{onS}), (\ref{onS'}) and (\ref{onH}), 
that on $H$
\be \nabla^{(n-2)}_{(A}\omega^{\ell'}_{B)}\ +\  \omega^{\ell'}_{A}\omega^{\ell'}_{B} \ -\ 
\frac{1}{2}R^{(n-2)}_{AB} \ +\ \frac{1}{2}R_{AB} \ =\ 0. \ee
Hence,  due to (\ref{LlSAB}),
\be {\cal L}_{\ell'} S'_{AB}\ =\ 0 \ee
on $H$.

Next, consider the generic case
\be \kappa^{\ell'}\ \not=\ 0.\ee
Similarly as we determined $\omega^{\ell'}_A$ at the slice $S$, we can also relate $S'_{AB}$ to $S_{AB}$
thereon. We have 
\be S'_{ab}\ =\ -\nabla_a\nabla_b v'\ =\ -\nabla_a\Big(\frac{1}{f}\nabla_b v\Big)\ =  \frac{1}{f}S_{ab} + \frac{1}{f^2}\nabla_a f \nabla_b v .\ee
Therefore, at $S$, 
\begin{eqnarray} S'_{AB}|_S\ =\ S_{AB}|_S \ =\ \frac{1}{\kappa^{\ell} }\left.\left( \nabla^{(n-2)}_{(A}\omega^{\ell}_{B)}\ +\  \omega^{\ell}_{A}\omega^{\ell}_{B} \ -\ \frac{1}{2}R^{(n-2)}_{AB} \ +\ \frac{1}{2}R_{AB}\right)\right|_S 
\end{eqnarray}
With that initial value $S'_{AB}$ is determined by the equation (\ref{LlSAB}). But notice, that a solution of (\ref{LlSAB}) with the same 
initial data is
\be\label{sol} S'_{AB}\ =\ \frac{1}{\kappa^{\ell'} }\left( \nabla^{(n-2)}_{(A}\omega^{\ell'}_{B)}\ +\  \omega^{\ell'}_{A}\omega^{\ell'}_{B} \ -\ 
\frac{1}{2}R^{(n-2)}_{AB} \ +\ \frac{1}{2}R_{AB}\right)   ,\ee
hence, by the uniqueness, this is it. 
The solution satisfies
\be {\cal L}_{\ell'} S'_{AB}\ =\ 0 .\ee
 That completes the proof.

\section{Composing spacetimes from horizons}

Suppose spacetime $M$ is foliated by non-expanding 
horizons.  A characterization of such spacetimes and horizons  is the goal of the current paper. 
\subsubsection{A distinguished null vector field}
Our first observation is that the existence of the foliation  allows us to distinguish a null vector field 
at each of the horizons up to rescaling by a  constant. Let 
 \be \label{u} u:M\rightarrow \mathbb{R}\ee
be a function such that   each non-expanding horizon - a lief of the foliation -   consists of the points of $M$ 
defined by the equation    
\be u(x^\alpha) \ =\ u_0.\ee
We denote the corresponding horizon  by $H_{u_0}$. Consider the vector field
\be\label{elldu} \ell^\alpha\ =\ -g^{\alpha\beta}\nabla_\beta u    \ee
(the minus sign is a convention only, for the consistency with the literature on the exact results to Einstein's equations 
we will refer to in the next section).  The vector field $\ell^\alpha$  is tangent to each of the horizons $H_u$.  
It satisfies
\be \ell^\alpha\ell_\alpha\ =\ 0, \ \ \ \ \ \ \ \ \ \nabla_{[\alpha}\ell_{\beta]}\ =\ 0 \ee 
hence it is also geodesic
\be \ell^\alpha\nabla_\alpha \ell^\beta\ =\ 0 .\ee
From the horizons point of view, that means that  the surface gravity vanishes
\be \label{kappa0} \kappa^\ell\ =\ 0 . \ee
Therefore,   owing  to (\ref{kappa0}) and (\ref{kappa})-(\ref{dkappa}) 
the corresponding 1-form potential $\omega^\ell$ satisfies, similarly to (\ref{l})-(\ref{Llg}),
\be  {\cal L}_\ell\omega^\ell_a\ =\ 0\ =\ \ell^a\omega^\ell_a ,\ee
on every $H_u$.

The function $u$ and the field $\ell$ are defined up to the following transformations
\be \tilde{u}\ =\ f(u),\ee
and, respectively,
\be \tilde{\ell}\ =\ f'(u)\ell . \ee
The factor $f'(u)$ is constant on each $H_u$, therefore indeed, at each horizon $H_u$, the vector field  $\ell$ is 
defined up to rescaling by a constant factor.  

\subsubsection{A new  constraint implied by the existence of a
  foliation}

As in the previous section we introduce a variable
\be v : M\rightarrow \mathbb{R}\ee
such that
\be \ell^\alpha \nabla_\alpha v\ =\ 1 . \ee 
The  spacial sections
\be v={\rm const} \ee
of horizons $H_{u}$ will be denoted by $S_{v}(u)$.    

The existence of the foliation consisting of the non-expanding horizons $H_u$  leads
to additional to (\ref{LlSab}) constraints on the intrinsic-extrinsic
geometry $(g_{ab},\nabla_c)$  
of each of  the horizons $H_{u}$.  We will present the derivation in
the next section.  Here we state  
the result and  discuss its geometric consequences.  
The new constraint is:  on each of the slices $S_v(u)$ the induced
metric tensor $g_{AB}$ and  the  
pullback $\omega^\ell_A$ of the rotation  1-form potential determine
the pullback  $R_{AB}$ of 
the Ricci tensor $R_{\alpha\beta}$  of the spacetime metric $g_{\alpha\beta}$  in the following way
\be\label{Rn}
R_{AB}\ =\  \nabla^{(n-2)}_{A}\omega^\ell_{B}\ +\ \nabla^{(n-2)}_{A}\omega^\ell_{B}\ -\  
2\omega^\ell_A\omega^\ell_B\ +\ R^{(n-2)}_{AB} . \ee    
On each $H_u$, the equation   (\ref{Rn}) holds for every value of
$v$. The right-hand-side of (\ref{Rn}) is  
Lie dragged  by the vector field $\ell$. It follows, that so is the left hand side
\be {\cal L}_\ell R_{AB} \ =\ 0 ,\ee 
and in the consequence of the first equation (\ref{RatH}) the
consistency condition on the spacetime Ricci  
tensor (that is on the energy momentum tensor $T_{\alpha\beta}$)  is
that its pullback $R_{ab}$ to each of  
the horizons $H_u$  satisfies
 \be\label{LlRab} {\cal L}_\ell R_{ab} \ =\ 0 .\ee
 Given $H_u$, one of the horizons, the equations (\ref{Rn}) induced on each spacelike  $n-2$ dimensional
 slice all amount to a single equation. In other words, when the consistency condition  (\ref{LlRab}) is satisfied
 on a horizon, and    (\ref{Rn})  is satisfied on a single  slice $S_v(u)$, the it is satisfied on every other slice $S_{v+f}(u)$, where $f$ is an arbitrary function. In view of the equation (\ref{Rn}) and (\ref{kappa0}), the Einstein constraint (\ref{LlSAB}) turns into
 \be\label{FolCond} {\cal L}_\ell S_{AB}\ =\   2\nabla^{(n-2)}_{(A}\omega^\ell_{B)}  .
\ee
 Notice, that the above condition involves only the geometry $(g_{AB},\nabla_C)$ of a given 
 horizon $H_u$, and is independent of $T_{\alpha\beta}$ which normally enters through $R_{AB}$. 

\subsubsection{Comparison with the extremal isolated horizon constraint}
Let us compare the very condition (\ref{Rn}) with the condition (\ref{Iso1}) on extremal isolated horizon.
At first glance they look similar. Certainly they coincide in the case
\be \omega^\ell_A\ =\ 0. \ee
More generally, they are related with each other by the transformation   
\be\label{wto-w} \omega^{\ell'}_A\ \mapsto\ -\omega^\ell_A .\ee
If we tried to achieve it within a same horizon, then possible transformations would be 
\be\label{...} \ell'\ =\ f\ell,\ \ \ \ \ \ \ \ \ell^a\nabla_a\ln f\ =\ 0, \ee
accompanied by
\be  \omega^{\ell'}_A\ =\ \omega^\ell_A + \nabla_A \ln f \ee
then, (\ref{wto-w}) implies
\be \omega^\ell_A\ =\ -\frac{1}{2}\nabla_A\ln f. \ee
That is, the function $f$ exists, provided $\omega^\ell_A$ is
itself a gradient. This is a severe restriction to a non-rotating 
\be \Omega_{AB}\ =\ 0  \ee  
extremal horizon case.         
   
\subsubsection{Bifurcated non-expanding horizons}
There is another geometric mechanism that implies the transformation (\ref{wto-w})
for a general non-expanding horizon, however, in a special spacetime.
 
Consider two intersecting non-expanding horizons $H$ and $H'$.  Choose null vectors
$\ell$ and $\ell'$ on $H$ and, respectively, $H'$ such that
\be \ell^\alpha\ell'_\alpha\ =\ -1 . \ee      
Then, on $H\cap H'$
\be\label{HintH'} \omega^{\ell'}_A\ =\ -\ell_\alpha
\nabla_A{\ell'}^\alpha\ =\ \ell'_\alpha
\nabla_A{\ell}^\alpha\ =\ -\omega^\ell_{A} \ee 

\subsubsection{A transversal non-expanding horizon}\label{sec:prop}
Combining (\ref{Rn}) transformed by (\ref{HintH'})   with the existence condition (\ref{Iso1}) for an extremal horizon, we conclude the following result:
\bigskip

{\nonumber{\bf Proposition}}
{\it Suppose spacetime $M$ is foliated by non-expanding horizons. Suppose, there exists
a transversal non-expanding horizon $H'$ and a lief $H$ of the foliation, such   $H'\cap H$ is a spacelike
section of each of them.   If  Einstein's equations (\ref{Ein}), and the energy conditions (\ref{energy}) are satisfied, then $H'$ admits 
an extremal isolated horizon structure defined in Sec.\ref{sec:iso}.  }

\section{Direct derivation} 
In this section we will derive the equality (\ref{Rn}) along with other components of the Ricci tensor.   

\subsubsection{Adapted coordinates}
Consider spacetime $M$ foliated by non-expanding horizons.  We will be using the 
functions $u$ and $v$ defined above in (\ref{u}), (\ref{v}), and the
vector field $\ell$ given in (\ref{elldu}).  
Let  
$$x^A,\ \ \ \ \ \ \ A=1,...,n-2$$ 
by additional functions  such that
\be \ell^\alpha \partial_\alpha x^A\ =\ 0. \ee
In this way we have obtained coordinates $(x^\alpha)=(x^A,v,u)$ referred to as adapted.
They are defined up to the following elementary transformations
\begin{align} 
   u &= f(u'),  &v &= \frac{v'}{f'(u')}, &x'^A &= x^A, \nonumber\\
   u &= u',     &v &= v' + f(x^A,u),       &x'^A &= x^A,    \label{trans}\\     
   u &= u',     &v &= v',                &x'^A &= f^A(x^B,u) \nonumber    
\end{align} 
and their compositions.

\subsubsection{Topological assumptions}
We are assuming the topology of $M$ to be
\be M\ =\ S\times\mathbb{R}\times\mathbb{R} \ee
where the surfaces 
$$u\ =\ {\rm const},$$
have the topology
\be H_u\ =\ S\times\mathbb{R},\ee
the surfaces 
$$u\ =\ {\rm const},\ \ \ v\ =\ {\rm const}$$
have the topology
\be S_v(u)\ =\ S\ee
and finally the surfaces $v={\rm const}$ have the topology of $S\times\mathbb{R}$.

\subsubsection{The spacetime and horizons geometry}
In the adapted coordinates $(x^\alpha)\ =\ (x^A,v,u)$, the metric takes the following form
\begin{align} 
g_{\alpha\beta}dx^\alpha dx^\beta\ &=\ g_{AB}dx^A dx^B\ -2du\left(dv + W_Adx^A+Hdu\right), \\
g_{AB,v}\ &=\ 0.
\end{align} 
The elements of the horizon structures introduced above are expressed in terms of
the adapted coordinates as follows
\begin{align}
\ell\ &=\ \frac{\partial}{\partial v}, \\
\omega^\ell\ &=\ \frac{1}{2}W_{A,v}dx^A,\\
\kappa^\ell\ &=\ 0,\\
S_{AB}\ &=\ \nabla^{(n-2)}_{(A}W_{B)} + \frac{1}{2}g_{AB,u}
\end{align}
where $\nabla^{(n-2)}_{(A}$ is the covariant derivative defined on each slice $u={\rm const}$ and $v={\rm const}$ 
by the metric $g_{AB}(x^C,u)$.

\subsubsection{The horizon constraints}
The zeroth low (\ref{dkappa}) implies
\be \label{zerothW}
W_{A,vv}\ =\ 0   .
\ee 
while the horizon constraint  (\ref{LlSAB}) reduces to (\ref{Rn}), because the condition (\ref{FolCond}) is satisfied identically by the very $S_{AB}$.
%

\subsubsection{The Ricci tensor components $R_{AB}$ and $R_{Av}$}
Calculation of  $R_{AB}$ gives 
\be \label{RAB}
    R_{AB}\ = \nabla^{(n-2)}_{(A}W_{B),v} - \frac{1}{2} W_{A,v}W_{B,v} + R^{(n-2)}_{AB} .
\ee
This result proves (\ref{Rn}). 
The $(A,v)$ component of the spacetime Ricci tensor   is 
 \be
    R_{Av}=\frac{1}{2}W_{A,vv}.
 \ee
 This component was already mentioned in (\ref{RatH}) in the context of a single horizon, and it was 
 assumed to vanish thereon as a consequence of the Einstein equations and the energy inequalities. 
 Since the horizons actually cover  the spacetime in  this case, we are assuming the vanishing everywhere
\be
    R_{Av}\ =\ 0 
 \ee
that is consistent with (\ref{zerothW}).

\subsubsection{The Ricci tensor component $R_{uv}$}
We also obtain new relations between the isolated horizons and the Ricci tensor, that were not mentioned in the previous section. 
One of them is quite simple and clearly determines the function $H_{,vv}$  by  the 1-form $\omega^\ell_A$ modulo the spacetime Ricci tensor, namely
\be
  R_{uv}   = -\frac{1}{2}\partial_v\left[\nabla^{(n-2)}_A
      W^A -W^A W_{A,v} - 2H_{,v} \right].   
\ee

\subsubsection{The Ricci tensor component $R_{Au}$ and $R_{uu}$}
The remaining equations are more complicated, namely
\begin{align}
    R_{Au} &= g^{BC}\left(\nabla^{(n-2)}_{[B}g_{A]C,u}+\nabla^{(n-2)}_B\nabla^{(n-2)}_{[C} W_{A]}\right)-
          \frac{1}{2}S^B{}_B W_{A,v} - \frac{1}{2}W_{A,uv} \nonumber\\
           &\quad - \frac{1}{2}W^B \nabla^{(n-2)}_B W_{A,v}
             +W^B \nabla^{(n-2)}_{[A} W_{B],v}+\frac{1}{2}W^B{}_{,v}\nabla^{(n-2)}_A W_B\\
           &\quad +\frac{1}{2}W^B
             W_{B,v}W_{A,v}+\left(H+\frac{1}{2}W^B W_B\right)W_{A,vv}
             + H_{,Av}, \nonumber \\ 
    R_{uu} &=  -g^{AB}\left(\nabla^{(n-2)}_A W_{B,u}+\frac{1}{2}g_{AB,uu}\right) + L^{AB}L_{AB}+
  2W^AW^B{}_{,v}\nabla^{(n-2)}_{[A}W_{B]} \nonumber \\ 
           &\quad  +W^AW_{A,uv}-\frac{1}{2}W^AW^BW_{A,v}W_{B,v}+
            (2H+W^AW_A)\big(H_{,vv}+\frac{1}{2}W^B{}_{,v}W_{B,v}\big) \\ 
           &\quad + W^A{}_{,v}H_{,A} - 2W^AH_{,Av} + g^{AB}\nabla^{(n-2)}_A H_{,B} -
            S^A{}_AH_{,v}, \nonumber       
\end{align}
where
 $$ L_{AB}=\nabla^{(n-2)}_{[A}W_{B]} + \frac{1}{2}g_{AB,u} $$ 
and they farther constraint the functions $W_A$ and $H$ modulo the spaceime Ricci tensor.




\section{The case of a transversal horizon}

In this section we follow the clue provided by Proposition. We
assume that in addition to the non-expanding horizon foliation, there exists in
the spacetime $M$ a transversal extremal horizon $H'$ (we denote
the null symmetry generator $\ell'$).  In the vacuum case we solve the Einstein's
equations completely  given arbitrary data on the extremal horizon: the rotation
1-form potential and the $n-2$-metric tensor on its spacelike slice subject
to the extremal isolated horizon constraints. 
  
\subsubsection{The general $R_{\alpha\beta}$ case}
Given $(H', \ell')$,  we farther adapt our coordinates $(x^A,v,u)$. 
Using the transformations (\ref{trans}),    we may adjust the coordinate $v$ 
such that the extremal horizon $H'$ is the surface
\be v\ =\ 0 .\ee
We may also adjust the coordinate $u$, such that  
\be \ell'^\alpha \partial_\alpha u\ =\ 1.\ee
Finally, we adjust the coordinates $x^A$ such  that
\be \ell'^\alpha\partial_\alpha x^A\ =\ 0 . \ee
In those coordinates,  
\be \ell'\ =\ \frac{\partial}{\partial u} . \ee 
The coordinate transformations (\ref{trans}) are now reduced to
\begin{align}  
              u & = f_0 u',  &v & =\frac{v'}{f_0}, &x'^A &= x^A\\
              u & = u',      &v & = v',            &x'^A & = f^A(x^B),
                       \label{transiso}
\end{align} 
where $f_0\ =\ {\rm const}$. 
The rotation 1-form potential of $H'$ is
 \be\label{omega'} \omega^{\ell'}_A(x,u)\ =\ -\frac{1}{2}W_{A,v}(x,u),\ee
where we denoted $x:=(x^A)$.
It follows from the non-expanding horizon properties of $H'$ that
\begin{align}   g_{AB,u}\ =\ 0,\\
W_A(x,v=0,u)\ =\ 0,\\
 H(x,v=0,u)\ =\ 0.
\end{align}
It follows from the isolated horizon assumption on $H'$ (the $0$ law) that
\be 
W_{A,vu}(x,v=0,u)\ =\ 0.
\ee 
Therefore we have,
\be W_A(x,v,u)\ =\  vW_{1A}(x) .  \ee
Finally, from the extremality
\be \kappa^{\ell'}\ =\ 0\ee
it follows that that 
\be H_{,v}(x,v=0,u)\ =\ \frac{1}{2}\ell^\alpha(\ell'^\beta\ell'_\beta)_{,\alpha}\ =\ \kappa^{\ell'}\ =\ 0 . \ee

\subsubsection{The vacuum case}
Suppose in addition to the assumptions of the previous subsection, 
that $g_{\alpha\beta}$ satisfies the vacuum equations
\be R_{\alpha\beta}\ =\ 0 .\ee
In this case 
\be 
  H_{,vv}\ =\  \frac{1}{2}\Big( \nabla^{(n-2)}_A W_1^A-W_1^A W_{1A} \Big).
\ee
Hence, using the rotation 1-form potential (\ref{omega'}) and  dropping the superscript
$$ \omega_{A}\ :=\ \omega_{A}^{\ell'} $$
 a general solution takes the form 
\be\label{solution} g_{\alpha\beta}dx^\alpha
dx^\beta\ =\ g_{AB}(x)dx^Adx^B - 2du\Big(dv  -2v\omega_Adx^A - \frac{1}{2}v^2\Big(
\nabla^{(n-2)}_A \omega^A + 2\omega^A \omega_{A} \Big)du\Big) \ee 
and is uniquely defined by a given solution $(g_{AB},\omega_{A})$ to the equation
(\ref{RAB})
\be \label{gomega} \nabla^{(n-2)}_{(A}\omega_{B)}\ +\  \omega_{A}\omega_{B} \ -\ 
\frac{1}{2}R^{(n-2)}_{AB}  =\ 0   \ee
defined on the $n-2$ manifold $S$ in terms of unknown metric tensor $g_{AB}$ and $1$-form
$\omega_A$.

The solution we have obtained belongs to a more general class of  metric tensors
\be g_{\alpha\beta}dx^\alpha dx^\beta\ =\ g_{AB}(x)dx^Adx^B - 
              2du\big(dv  -2v\omega_A(x)dx^A + v^2H_2(x)du\big) \ee
where $g_{AB}(x),\, \omega_A(x)$ and $H_2(x) $ are arbitrary.  This class of metrics is known
as describing near horizon geometries \cite{LivRevNHG}. Each of them has the following two Killing vector fields
\be \xi_0\ =\ \partial_u \ee
and 
\be \xi_1\ =\ v\partial_v - u\partial_u.\ee
The surface 
$$v=0$$ 
is an extremal Killing horizon of the Killing vector $\xi_0$.  At the same time, this surface
is a part of the bifurcated non-extremal  Killing horizon of a  Killing vector field 
\be K_{u_0}\ =\ u_0\xi_0 \ +\ \xi_1,  \ee
the second part being the surface
\be\label{u_0} u=u_0.\ee
Therefore, our  solution (\ref{solution}) as well as the general near horizon metric 
admit  foliation 
$$ u\ =\ {\rm const}$$
defined by the Killing horizons.  Equivalently to our derivation, the solution 
(\ref{solution}), (\ref{gomega}) 
can be obtained by imposing the vacuum Einstein's equations  on near horizon geometries (\cite{LivRevNHG}).

\section{Summary} We considered spacetime foliated by non-expanding horizons. 
A necessary existence condition is a  constraint (\ref{FolCond}) on the geometry satisfied by each of  the 
horizons, which is intrinsic  in the sense that  it does not involve the stress energy tensor. 
 We have also derived a constraint  (\ref{Rn}) between the geometry of each horizon and 
 the stress energy tensor. The constraint has an interesting form: it can be mapped  into the extremal isolated 
 horizon constraint (\ref{Iso1})  by the transformation (\ref{wto-w}).  Surprisingly,  the transformation relates
the rotation 1-form potential of each horizon with the 1-form potential of a transversal
extremal horizon. The exact sense of that interpretation is provided by Proposition (see Sec. \ref{sec:prop}), 
which states, that if there exists one more non-expanding horizon $H'$, transversal to the foliation, then its rotation 1-form 
potential is defined by the transformation (\ref{wto-w}), and  the condition (\ref{Rn}) ensures the 
existence on $H'$ of the extremal isolated horizon structure. 

We derived our results by introducing  coordinates adapted to the foliation and imposing Einstein's   equations.  
In the case of the existence of an additional transversal non-expanding horizon,  we were able to solve the vacuum 
equations completely for every given solution of the equation (\ref{gomega}).  The solution is the metric tensor (\ref{solution}). 
It has the form known also as near horizon geometry. A transversal horizon free case  is more mysterious.

 \section*{Acknowledgments}
This work was partially supported by the Polish National
Science Centre grant No.~2015/17/B/ST2/02871.



\begin{thebibliography}{99}

\bibitem{PLJ} T. Paw\l owski, J. Lewandowski, J. Jezierski,
\emph{Spacetimes foliated by Killing horizons},
Class. Quant. Grav. {\bf 21} (2004), 1237--1252, arXiv:gr-qc/0306107.

\bibitem{ABL}A. Ashtekar, C. Beetle, J. Lewandowski,
\emph{Geometry of Generic Isolated Horizon},
Class. Quant. Grav. {\bf 19} (2002), 1195--1225, arXiv:gr-qc/0111067.

\bibitem{LPuniq}J. Lewandowski, T. Paw\l owski,
\emph{Extremal Isolated Horizons: A Local Uniqueness Theorem},
Class. Quant. Grav. {\bf 20} (2003), 587--606, arXiv:gr-qc/0208032.

\bibitem{Horowitz}J. M. Bardeen, G. T. Horowitz,
\emph{The Extreme Kerr throat geometry: A Vacuum analog of $AdS(2)\times S^2$},
Phys. Rev. D {\bf 60} (1999), 104030, arXiv:hep-th/9905099.


\bibitem{ExactSolutions} H. Stephani, D. Kramer, M. MacCallum, C. Hoenselaers, E. Herlt,
\emph{Exact Solutions to Einstein's Field Equations},
Cambridge University Press, 2003.
 
\bibitem{Jez}J. Jezierski,
\emph{On the existence of Kundt's metrics and degenerate (or extremal) Killing horizons},
Class. Quant. Grav. {\bf 26} (2009), 035011, arXiv:0806.0518 [gr-qc].

\bibitem{JezKerr}J. Jezierski, B. Kaminski,
\emph{Towards uniqueness of degenerate axially symmetric Killing horizon},
Gen. Relativ. Gravit. {\bf 45} (2016), 987--1004, arXiv:1206.5136 [gr-qc].

\bibitem{LivRevNHG} H. K. Kunduri, J. Lucietti,
\emph{Classification of Near-Horizon Geometries of Extremal Black Holes},
Living Rev. Rel. {\bf 16} (2013), 8, http://www.livingreviews.org/lrr-2013-8, arXiv:abs/1306.2517.

\bibitem{Reall} H. S. Reall,
\emph{Higher dimensional black holes and supersymmetry},
Phys. Rev. D {\bf 68} (2003), 024024, arXiv:hep-th/0211290.

\bibitem{LPhigh}J. Lewandowski, T. Paw\l owski,
\emph{Quasi-local rotating black holes in higher dimension: geometry},
Class. Quant. Grav. {\bf 22} (2005), 1573--1598, arXiv:gr-qc/0410146.

\bibitem{LivRevIH}A. Ashtekar, B. Krishnan,
\emph{Isolated and Dynamical Horizons and Their Applications},
Living Rev. Rel. {\bf 7} (2004), 10, http://www.livingreviews.org/lrr-2004-10, arXiv:gr-qc/0407042.

\end{thebibliography}
\end{document}